\documentclass[aps, prd, onecolumn, tightenlines, notitlepage, superscriptaddress, nofootinbib, preprintnumbers, floatfix,showkeys,11pt]{revtex4-2}
\usepackage[normalem]{ulem}
\usepackage{amstext}
\usepackage{amssymb}
\usepackage{amsmath}
\usepackage{graphicx}
\graphicspath{{plots/}}
\usepackage{url}
\usepackage{color}
\usepackage{caption}
\usepackage{subcaption}
\usepackage{placeins} 
\usepackage{ulem}
\usepackage[utf8]{inputenc}
\pdfoutput=1
\usepackage{textcomp}
\usepackage{comment}
\usepackage{yfonts}
\usepackage{epsfig,amsfonts,mathrsfs,amsmath,amssymb,graphicx,color,slashed,multirow}
\usepackage{amsmath,latexsym,amssymb,graphicx,slashed,color,enumerate,url,cancel,gensymb}
\usepackage{textcomp}

\usepackage[x11names]{xcolor}
\usepackage[colorlinks]{hyperref}

\usepackage{booktabs}
\usepackage{adjustbox}

\definecolor{vdrgreen}{rgb}{0.0, 0.7, 0.0}

\def\cevns{CE$\nu$NS }
\def\eves{E$\nu$ES }

\definecolor{indianred}{rgb}{0.8, 0.36, 0.36}
\definecolor{blue(ncs)}{rgb}{0.0, 0.53, 0.74}

\AtBeginDocument{\hypersetup{citecolor=indianred,linkcolor=indianred,urlcolor=indianred}}

\usepackage{float}

\usepackage[T1]{fontenc}
\usepackage{ae,aecompl}
\graphicspath{{Figures/}}
\usepackage{appendix}



\newcommand{\AddrIISERB}{Department of Physics, Indian Institute of Science Education and Research - Bhopal, \\ 
Bhopal Bypass Road, Bhauri, Bhopal 462066, India}

\usepackage{orcidlink}
\begin{document}

\title{\LARGE \textcolor{indianred}{Probing Standard Model and Beyond with Reactor CE$\nu$NS Data of CONUS+ experiment}}
\author{Ayan Chattaraj~\orcidlink{0009-0001-3561-7049}}
\email{ayan23@iiserb.ac.in}
\affiliation{\AddrIISERB}
\author{Anirban Majumdar~\orcidlink{0000-0002-1229-7951}}\email{anirban19@iiserb.ac.in}
\affiliation{\AddrIISERB}
\author{Rahul Srivastava~\orcidlink{0000-0001-7023-5727}}
\email{rahul@iiserb.ac.in}
\affiliation{\AddrIISERB}

\begin{abstract}
\vspace{1cm}
We explore the potential of reactor antineutrino-induced Coherent Elastic Neutrino-Nucleus Scattering (CE$\nu$NS) data from the CONUS+ experiment to investigate both the Standard Model (SM) and Beyond Standard Model (BSM) scenarios. Alongside CE$\nu$NS, Elastic Neutrino-Electron Scattering (E$\nu$ES) events are also included in our analysis, enabling more stringent constraints on new physics. Within the SM, we examine the weak mixing angle as a precision test of the electroweak sector. For BSM scenarios, we constrain the parameter space of light mediators arising from neutrino generalized interactions (NGI), while also setting limits on the electromagnetic properties of neutrinos, including their charge radius, millicharge, and magnetic moment. 
\end{abstract}
\maketitle

\section{Introduction}
\label{sec:intro}

Coherent Elastic Neutrino-Nucleus Scattering (CE$\nu$NS) is a neutral current process within the Standard Model (SM) mediated by the exchange of a $Z$ boson. In this process, a neutrino elastically scatters off an entire nucleus, resulting in a nuclear recoil while the nucleus remains in its ground state. Furthermore, the coherence condition is satisfied when the momentum transfer is smaller than the inverse of the nuclear radius. Although CE$\nu$NS was first predicted by Freedman in 1974~\cite{Freedman:1973yd}, its experimental detection was challenging due to the small nuclear recoil energies involved. The first observation of CE$\nu$NS was achieved more than four decades later in 2017 by the COHERENT collaboration at Oak Ridge National Laboratory, USA, using a CsI detector~\cite{COHERENT:2017ipa}. Subsequent detections with liquid argon (LAr)~\cite{COHERENT:2020iec} and germanium (Ge)~\cite{COHERENT:2024axu} detectors in 2020 and 2024, respectively, confirmed this initial observation.

In 2022, the Dresden-II collaboration reported CE$\nu$NS detection using reactor antineutrinos, marking a significant milestone for reactor-based experiments~\cite{Colaresi:2022obx}. Recently, the CONUS+ collaboration independently observed CE$\nu$NS using reactor antineutrinos~\cite{Ackermann:2025obx}. Reactor-based sources are particularly well-suited for studying CE$\nu$NS due to their low-energy neutrino spectrum, which ensures a near-unity nuclear form factor and a maximally coherent process.

The CONUS+ experiment utilized a high antineutrino flux of $1.5 \times 10^{13}~\mathrm{cm^{-2}s^{-1}}$ provided by the boiling water reactor at the Leibstadt Nuclear Power Plant (KKL) in Switzerland. The reactor operates at a thermal power of 3.7~GW, and the detector array, consisting of three high-purity germanium (HPGe) detectors (C2, C3, and C5) with fiducial masses of 0.95 kg, 0.94 kg, and 0.94 kg, respectively, was positioned at 20.7 meters from the reactor core. These detectors had energy thresholds of $T_e^\mathrm{th} = 160 \mathrm{~eV}_{ee}$ for C3, $T_e^\mathrm{th} = 170 \mathrm{~eV}_{ee}$ for C5, and $T_e^\mathrm{th} = 180 \mathrm{~eV}_{ee}$ for C2. The CE$\nu$NS measurement was performed over different reactor-on periods for each detector (117 days for C2, 110 days for C3, and 119 days for C5), leading to a total fiducial exposure of $327 \mathrm{~kg} \times \mathrm{days}$. The collaboration detected $395 \pm 106$ CE$\nu$NS events, achieving a $3.7\sigma$ confidence level of CE$\nu$NS detection.

In this paper, we explore the potential of CE$\nu$NS data from the CONUS+ experiment to probe both SM and Beyond Standard Model (BSM) physics. Elastic Neutrino-Electron Scattering (E$\nu$ES), a concurrent process to CE$\nu$NS, also contributes to the observed signals. Within the SM, the E$\nu$ES contribution to the total event rate is negligible at low recoil energies and is therefore often disregarded in CE$\nu$NS analyses. However, certain BSM scenarios can significantly enhance the E$\nu$ES contribution, making its inclusion essential for deriving stronger constraints on BSM parameters. Consequently, in this work, we incorporate E$\nu$ES signals alongside CE$\nu$NS signals in all cases to ensure a comprehensive analysis. Within the SM, we determine the weak mixing angle as a test of electroweak precision. Concerning BSM interactions, we investigate signatures of new mediators arising from neutrino generalized interactions (NGI) and place constraints on the electromagnetic properties of neutrinos, including their charge radius, millicharge, and magnetic moment.

The remainder of this paper is organized as follows. In Sec.~\ref{Sec:Theory}, we describe the theoretical framework and present the CE$\nu$NS and E$\nu$ES cross sections within the SM and under different BSM scenarios. Sec.~\ref{Sec:Data_Analysis} details the methodology used for event simulation and statistical analysis, including the experimental specifications of the CONUS+ experiment. In Sec.~\ref{Sec:Results}, we present and discuss our results, highlighting the sensitivity of the CONUS+ experiment to SM and BSM phenomena. Finally, in Sec.~\ref{Sec:Conclusions}, we summarize our findings and discuss their implications.


\section{\label{Sec:Theory} Theoretical framework}

\subsection{\cevns and \eves within the Standard Model Framework}

In the SM, the differential cross section for \cevns arises from a $t$-channel neutral current interaction. At low  neutrino energies ($E_\nu \ll M_Z$), the differential cross section is expressed as~\cite{Freedman:1973yd, Drukier:1984vhf}:
\begin{equation}
\label{Eq:CEvNS_SM_xSec}
\begin{split}
\left[\frac{\mathrm{d}\sigma}{\mathrm{d}T_\mathcal{N}}\right]_{\mathrm{SM}}^{\mathrm{CE}\nu\mathrm{NS}} = \frac{G_F^2 m_\mathcal{N}}{\pi} &\mathcal{F}_W^2(\mathfrak{q}^2) \, (Q_{V}^\mathrm{SM})^2 \left(1 - \frac{m_\mathcal{N} T_\mathcal{N}}{2 E_{\nu}^2}\right) \, ,
\end{split}
\end{equation}
where $G_F$ is the Fermi constant, $m_\mathcal{N}$ is the nuclear mass, $T_\mathcal{N}$ is the nuclear recoil energy, and $\mathcal{F}_W(\mathfrak{q}^2)$ represents the nuclear form factor at momentum transfer $\mathfrak{q}=\sqrt{2m_\mathcal{N}T_\mathcal{N}}$. In our present work we have considered the nuclear effect using a Helm type effective form factor~\cite{PhysRev.104.1466}. The SM weak vector charge, $Q_{V}^\mathrm{SM}$, is given by:
\begin{equation}
Q_{V}^\mathrm{SM} = Z\left(\frac{1}{2} - 2\sin^2\theta_W\right) - \frac{N}{2}\,,
\label{eq:weak_charge_SM}
\end{equation}
where $Z$ and $N$ are the proton and neutron numbers in the nucleus, respectively, and $\sin^2\theta_W$ is the weak mixing angle.

At the $Z$-pole, the weak mixing angle is precisely measured as $\sin^2\theta_W(M_Z) = 0.23121 \pm 0.00004$. However, at lower energy scales relevant to CE$\nu$NS (i.e., $\mathfrak{q}^2 \to 0$), its value is less constrained experimentally. Theoretically, its value at low energy is determined using the Renormalization Group Equations (RGE). Within the $\overline{\text{MS}}$ scheme, it is predicted as $\sin^2\theta_W(\mathfrak{q} = 0) = 0.23857 \pm 0.00005$~\cite{ParticleDataGroup:2022pth, Erler:2019hds}. The $\sin^2\theta_W$ dependence makes \cevns a promising probe for testing the weak mixing angle at low momentum transfer. Variations in $\sin^2\theta_W$ around its predicted value cause fluctuations in both the \cevns cross section and the predicted event rate, offering a means to measure this parameter at low energy scales. While analyses of data from experiments like COHERENT (CsI and LAr) have provided constraints on $\sin^2\theta_W$ at specific energy scales~\cite{DeRomeri:2022twg}, reactor neutrino experiments such as CONUS+ can offer complementary measurements. As reactor experiments operate in a different energy regime, it presents a unique opportunity to probe the weak mixing angle at a different energy scale, thereby enhancing our understanding of electroweak interactions in the low-energy domain.

On the other hand within the SM, the tree-level differential cross section for \eves with respect to the electron recoil energy, $T_e$, is expressed as~\cite{Kayser:1979mj}:
\begin{widetext}
\begin{equation}
\label{equn:EvES_SM_xsec_for_nu-alpha}
\left[\frac{\mathrm{d}\sigma_{\nu_\ell}}{\mathrm{d}T_e}\right]_{\mathrm{SM}}^{\mathrm{E}\nu\mathrm{ES}} = \frac{G_F^2 m_e}{2\pi} \left[ (\textsl{g}_V \pm \textsl{g}_A)^2 + (\textsl{g}_V \mp \textsl{g}_A)^2 \left( 1 - \frac{T_e}{E_\nu} \right)^2 - (\textsl{g}_V^2 - \textsl{g}_A^2) \frac{m_e T_e}{E_\nu^2} \right]\,,
\end{equation}
\end{widetext}
where $m_e$ is the electron mass, and the $+$ ($-$) sign corresponds to neutrino (antineutrino) scattering, respectively. The parameters $\textsl{g}_V$ and $\textsl{g}_A$ represent the vector and axial-vector couplings, respectively, and are defined as:
\begin{equation}
\label{table:EvES_SM_couplings}
\textsl{g}_V = -\frac{1}{2} + 2\sin^2\theta_W + \delta_{\ell e}, \qquad \textsl{g}_A = -\frac{1}{2} + \delta_{\ell e}.
\end{equation}
Here, $\delta_{\ell e}$ is the Kronecker delta, which accounts for the presence of charged-current interactions in the cross section. This term applies only to $\nu_e$--$e^-$ and $\bar{\nu}_e$--$e^-$ scattering processes, where $\delta_{\ell e} = 1$. For all other neutrino flavors, $\delta_{\ell e} = 0$, as there are no charged-current contributions. 

\subsection{Neutrino Electromagnetic properties}

The observation of neutrino oscillations~\cite{McDonald:2016ixn, Kajita:2016cak}, which imply nonzero neutrino masses~\cite{Pontecorvo:1957cp,Maki:1962mu}, provides a strong motivation to explore the nontrivial electromagnetic (EM) properties of neutrinos~\cite{Schechter:1981hw, Nieves:1981zt, Kayser:1982br, Shrock:1982sc}. The most general EM vertex for neutrinos can be described using the EM form factors $F_q(\mathfrak{q}^2)$, $F_\mu(\mathfrak{q}^2)$, $F_\epsilon(\mathfrak{q}^2)$, and $F_a(\mathfrak{q}^2)$~\cite{Vogel:1989iv} (for a comprehensive review, see Ref.~\cite{Nowakowski:2004cv, Giunti:2014ixa, Giunti:2024gec}). At zero momentum transfer, these form factors correspond to the neutrino millicharge, magnetic dipole moment, electric dipole moment, and anapole moment, respectively. For neutrino scattering processes, it is convenient to define an effective magnetic moment $\mu_{\nu_\ell}^\mathrm{eff}$, expressed in terms of the fundamental neutrino magnetic ($\mu$) and electric ($\epsilon$) dipole moments. In the context of short-baseline experiments, such as CONUS+, $\mu_{\nu_\ell}^\mathrm{eff}$ can be written as~\cite{Grimus:1997aa, AristizabalSierra:2021fuc}:
\begin{equation}
    \mu_{\nu_\ell}^\mathrm{eff} = \sum_k \left| \sum_i U^*_{\ell k} \lambda_{jk} \right|^2 ,
\end{equation}
where $\lambda_{jk} = \mu_{jk} - i \epsilon_{jk}$ are the fundamental transition magnetic moments (TMM), and $U$ represents the neutrino mixing matrix.

Within the SM, neutrinos are electrically neutral and do not have a magnetic moment. The neutrino magnetic moment, $\mu_{\nu_\ell}^\mathrm{eff}$, typically arises in various extensions of the SM required to explain the neutrino masses and mixing. It is generally small, as it usually originates from radiative corrections~\cite{Petcov:1976ff, Marciano:1977wx, Lee:1977tib, Fujikawa:1980yx, Pal:1981rm, Shrock:1982sc, Dvornikov:2003js, Dvornikov:2004sj}. However, certain new physics scenarios can lead to significantly enhanced values of the neutrino magnetic moment and charge radius (CR), as well as the existence of a neutrino millicharge ($q_{\nu_{\ell\ell'}}$)~\cite{Babu:1989wn}. Thus, the detection of any nontrivial electromagnetic properties in neutrinos would indicate the presence of physics beyond the SM. Experimental measurements of these parameters are therefore crucial for testing the SM and probing new physics.

The helicity-flipping neutrino magnetic moment contribution to the CE$\nu$NS and E$\nu$ES cross sections adds incoherently to the SM contribution and is given by~\cite{Vogel:1989iv}:
\begin{subequations}
\begin{equation}
\label{equn:CEvNS_mag_xsec_for_nu-alpha}
\begin{aligned}
\left[\frac{\mathrm{d}\sigma_{\nu_\ell}}{\mathrm{d} T_\mathcal{N}}\right]_\text{mag}^{\mathrm{CE}\nu\mathrm{NS}} =& \frac{\pi \alpha_\text{EM}^2}{m_e^2} \left[ \frac{1}{T_\mathcal{N}} - \frac{1}{E_\nu} \right] Z^2 \mathcal{F}_W^2(\mathfrak{q}^2) \left( \frac{\mu^\text{eff}_{\nu_\ell}}{\mu_B} \right)^2 \,,
\end{aligned}
\end{equation}
\begin{equation}
\label{equn:EvES_mag_xsec_for_nu-alpha}
\begin{aligned}
\left[\frac{\mathrm{d}\sigma_{\nu_\ell}}{\mathrm{d} T_e}\right]_\text{mag}^{\mathrm{E}\nu\mathrm{ES}} =& \frac{\pi \alpha_\text{EM}^2}{m_e^2} \left[ \frac{1}{T_e} - \frac{1}{E_\nu} \right] \left( \frac{\mu^\text{eff}_{\nu_\ell}}{\mu_B} \right)^2 \,,
\end{aligned}
\end{equation}
\end{subequations}
where $\mu_B$ is the Bohr magneton, and $\alpha_\text{EM}$ is the fine structure constant.

The helicity-preserving EM contributions due to millicharge ($q_{\nu_{\ell\ell'}}$) and neutrino CR ($\langle r^2_{\nu_{\ell\ell'}} \rangle$) can be expressed using the following differential cross sections~\cite{Giunti:2014ixa, Giunti:2024gec}:
\begin{widetext}
\begin{subequations}
\label{Eq:millichare_CR}
\begin{equation}
\label{equn:CEvNS_HP_EM_xsec_for_nu-alpha}
\begin{aligned}
\left[\frac{\mathrm{d}\sigma_{\nu_\ell}}{\mathrm{d} T_\mathcal{N}}\right]_{\mathrm{SM}+\mathrm{EM}}^{\mathrm{CE}\nu\mathrm{NS}} =& \frac{G_F^2 m_\mathcal{N}}{\pi} \left(1 - \frac{m_\mathcal{N} T_\mathcal{N}}{2 E_{\nu}^2}\right) \mathcal{F}_W^2(\mathfrak{q}^2) \left\{ \left[ \left( \frac{1}{2} - 2\sin^2{\theta_W} - Q_{\ell\ell} \right) Z - \frac{N}{2} \right]^2 + Z^2 \sum_{\ell' \neq \ell} \left| Q_{\ell\ell'} \right|^2 \right\} \,,
\end{aligned}
\end{equation}
\begin{equation}
\label{equn:EvES_HP_EM_xsec_for_nu-alpha}
\begin{aligned}
\left[\frac{\mathrm{d}\sigma_{\nu_\ell}}{\mathrm{d} T_e}\right]_{\mathrm{SM}+\mathrm{EM}}^{\mathrm{E}\nu\mathrm{ES}} = \frac{G_F^2 m_e}{2\pi} \Big\{ &\Big[(\textsl{g}_V \pm \textsl{g}_A+Q_{\ell\ell})^2+\sum_{\ell' \neq \ell}\left|Q_{\ell\ell'}\right|^2\Big] + \Big[(\textsl{g}_V \mp \textsl{g}_A+Q_{\ell\ell})^2+\sum_{\ell' \neq \ell}\left|Q_{\ell\ell'}\right|^2\Big] \left( 1 - \frac{T_e}{E_\nu} \right)^2\\
&- \Big[(\textsl{g}_V+Q_{\ell\ell})^2 - \textsl{g}_A^2+\sum_{\ell' \neq \ell}\left|Q_{\ell\ell'}\right|^2\Big] \frac{m_e T_e}{E_\nu^2} \Big\}\,,
\end{aligned}
\end{equation}
\end{subequations}
\end{widetext}
where $Q_{\ell\ell'} = \sqrt{2} \pi \alpha_\text{EM} / G_F \left( \langle r^2_{\nu_{\ell\ell'}} \rangle / 3 - 2 / \mathfrak{q}^2 \cdot (q_{\nu_{\ell\ell'}} / e) \right)$, with $e$ being the electron charge\footnote{Notice that the magnitude of the three momentum transfer for \cevns process is given as, $\mathfrak{q}=\sqrt{2m_\mathcal{N}T_\mathcal{N}}$, while for the \eves it is expressed as, $\mathfrak{q}=\sqrt{2m_eT_e}$.}. Here, $\langle r^2_{\nu_{\ell\ell}} \rangle$ and $q_{\nu_{\ell\ell}}$ represent the diagonal components of the neutrino CR and millicharge, while $\langle r^2_{\nu_{\ell\ell'}} \rangle$ and $q_{\nu_{\ell\ell'}}$ are the off-diagonal components, also referred to as neutrino transition CR and millicharge in the flavor basis.

\subsection{Neutrino Generalized Interactions}

In this section, we explore extensions of the SM by introducing light mediators. We take a model independent approach and consider all possible Lorentz-invariant bilinear interactions between neutrinos and quarks/electrons, commonly referred to as neutrino generalized interactions (NGIs). Within this framework, the most general Lagrangian can be written as~\cite{Lindner:2016wff, AristizabalSierra:2018eqm, Flores:2021kzl}:
\begin{equation}
\label{Eq:NGI_Lagrangian}
\mathscr{L}_\mathrm{NGI} \supset -\frac{G_F}{\sqrt{2}} \sum_{\substack{X=S,P,V,A,T\\f=u,d,e}} \varepsilon_{\nu f}^X\left[\bar{\nu} \Gamma^X  \nu \right] \left[\bar{f} \Gamma_X f \right] \, ,
\end{equation}
where $\Gamma_X \equiv \{\mathbb{I}, i\gamma^5, \gamma_\rho, \gamma_\rho\gamma^5, \sigma_{\rho\delta}\}$ with $\sigma_{\rho\delta} = \frac{i}{2}[\gamma_\rho, \gamma_\delta]$ corresponding to the scalar ($S$), pseudoscalar ($P$), vector ($V$), axial-vector ($A$), and tensor ($T$) interactions, respectively.

As the CONUS+ experiment utilizes a Ge detector, it is important to note that most of the Ge isotopes consist of even-even nuclei. For these spin 0 isotopes, the spin-dependent pseudoscalar, axial-vector, and tensor interactions vanish entirely~\cite{Chattaraj:2025rtj}. Only $^{73}$Ge, which has an odd neutron number and non-zero spin, can have non-vanishing contributions from the pseudoscalar, axial-vector, and tensor components~\cite{Chattaraj:2025rtj}. However, the abundance of $^{73}$Ge in natural Ge is very small ($\sim7.75\%$). As a result, these spin-dependent interactions contribute negligibly to the overall signal. 
This implies that CONUS+ data will not be able to put any meaningful  limits on P, A, T interactions.
Henceforth, we focus exclusively on spin-independent scalar and vector interactions.

We start with considering the vector interaction. For that we consider $U(1)_{B-L}$ model as a benchmark, which incorporates a new mediator field, $Z'$. The relevant part of the Lagrangian describing the interaction of $Z'$ with fermions in the context of \cevns and \eves processes is given by:
\begin{widetext}
\begin{equation}
    \mathscr{L}_{B-L} \supset \textsl{g}_{B-L} \left(Q^q_{B-L} \bar{q} \gamma^\mu q + Q^e_{B-L} \bar{e} \gamma^\mu e + Q^\nu_{B-L} \bar{\nu}_L \gamma^\mu \nu_L \right) Z'_\mu + \frac{1}{2} M_{Z'}^2 Z'_\mu Z'^\mu \,.
\end{equation}
\end{widetext}
Here, the charges of the fermions under the $U(1)_{B-L}$ symmetry are determined by anomaly cancellation conditions, resulting in $Q^q_{B-L} = 1/3$, and $Q^e_{B-L} = Q^\nu_{B-L} = -1$.  Within this framework, the \cevns differential cross section is modified as a rescaling of the SM cross section~\cite{Bertuzzo:2021opb, Majumdar:2024dms}:
\begin{widetext}
\begin{equation}
    \left[\frac{\mathrm{d}\sigma}{\mathrm{d}T_\mathcal{N}}\right]_{\mathrm{SM}+Z'}^{\mathrm{CE\nu NS}} = 
    \left(1 + \frac{Q_{Z'} \textsl{g}^2_{B-L}}{\sqrt{2} G_F Q_V^{\mathrm{SM}} \left(M_{Z'}^2 + 2 m_\mathcal{N} T_\mathcal{N}\right)}\right)^2 
    \left[\frac{\mathrm{d}\sigma}{\mathrm{d}T_\mathcal{N}}\right]_{\mathrm{SM}}^{\mathrm{CE\nu NS}} \,,
\end{equation}    
\end{widetext}
where the $Z'$-mediated effective nuclear charge is expressed as:
\begin{equation}
    Q_{Z'} = Q^\nu_{B-L} \left[ Z \left(2 Q^u_{B-L} + Q^d_{B-L}\right) + N \left(Q^u_{B-L} + 2 Q^d_{B-L}\right)\right] = -A
\end{equation}
Here, $A = Z + N$ is the mass number of the nucleus.  For E$\nu$ES, the differential cross section can be derived from the SM expression by modifying the vector coupling as~\cite{Lindner:2018kjo}:
\begin{widetext}
\begin{equation}
    \textsl{g}_{V} \rightarrow \textsl{g}'_{Z'} = \textsl{g}_{V} + \frac{\textsl{g}^2_{B-L} Q^\nu_{B-L} Q^e_{B-L}}{\sqrt{2} G_F \left(M_{Z'}^2 + 2 m_e T_e\right)} 
    = \textsl{g}_{V} + \frac{\textsl{g}^2_{B-L}}{\sqrt{2} G_F \left(M_{Z'}^2 + 2 m_e T_e\right)} \,.
\end{equation}
\end{widetext}

Next, we examine scenarios involving a possible light scalar mediator. We extend the SM by introducing a $CP$-even real scalar boson $\phi$ with mass $M_\phi$. The relevant part of the Lagrangian for \cevns and \eves processes is given by:
\begin{equation}
    \mathscr{L}_\phi \supset \left(\textsl{g}^q_\phi \bar{q} q + \textsl{g}^e_\phi \bar{e} e + \textsl{g}^\nu_\phi \bar{\nu}_R \nu_L \right) \phi - \frac{1}{2} M_\phi^2 \phi^2 \,.
\end{equation}
The scalar interaction creates an additive contribution to the SM cross section for both \cevns and E$\nu$ES. For CE$\nu$NS, the corresponding differential cross section is given as~\cite{Farzan:2018gtr},
\begin{equation}
    \left[\frac{\mathrm{d}\sigma}{\mathrm{d}T_\mathcal{N}}\right]_\phi^{\mathrm{CE\nu NS}} = 
    \frac{m_\mathcal{N}^2 T_\mathcal{N} Q_\phi^2}{4 \pi E_\nu^2 \left(M_\phi^2 + 2 m_\mathcal{N} T_\mathcal{N}\right)^2} \mathcal{F}_W^2(\mathfrak{q}^2) \,,
\end{equation}
where the effective nuclear coupling for scalar interactions is:
\begin{equation}
    Q_\phi = \textsl{g}^\nu_\phi \left(Z \sum_{q = u, d} \textsl{g}^q_\phi \frac{m_p}{m_q} f_{T_q}^p + N \sum_{q = u, d} \textsl{g}^q_\phi \frac{m_n}{m_q} f_{T_q}^n \right) \,.
\end{equation}
Here, $m_p$ and $m_n$ are the proton and neutron masses, $m_q$ represents the quark masses, and the hadronic structure parameters for scalar interactions are $f_{T_u}^p = 0.026 \,, \qquad f_{T_d}^p = 0.038 \,, \qquad f_{T_u}^n = 0.018 \,, \qquad f_{T_d}^n = 0.056$~\cite{DelNobile:2021wmp}.

The differential cross section for the scalar-mediated E$\nu$ES process can be written as~\cite{Link:2019pbm}:
\begin{equation}
    \left[\frac{\mathrm{d}\sigma_{\nu_\ell}}{\mathrm{d}T_e}\right]_\phi^{\mathrm{E\nu ES}} = 
    \frac{m_e^2 T_e \left(\textsl{g}^\nu_\phi \textsl{g}^e_\phi\right)^2}{4 \pi E_\nu^2 \left(M_\phi^2 + 2 m_e T_e\right)^2} \,.
\end{equation}
\section{Data Analysis}
\label{Sec:Data_Analysis}
In this section we provide an overview of the CONUS+ experiment~\cite{Ackermann:2025obx}, outlining its key specifications and describing the methods used to simulate the expected \cevns and \eves signals. Additionally, the statistical framework employed in this study is discussed in detail.

The expected number of theoretical \cevns events in the $i$th energy bin is calculated as follows:
\begin{widetext}
\begin{equation}
\label{equn:CEvNS_Events}
\left[R_i\right]_\xi^{\text{CE}\nu\text{NS}} = t_{\text{run}} N_{\text{target}} \int_{T_e^i}^{T_e^{i+1}} \hspace{-0.7cm}\mathrm{d}T_e^\mathrm{reco}\hspace{0.05cm}\int_{T_e^\mathrm{min}}^{T_e^\mathrm{max}}\hspace{-0.6cm}\mathrm{d}T_e~\mathcal{G}\left(T_e,T_e^\mathrm{reco}\right) \left(\frac{1}{Q_f} - \frac{T_e}{Q_f^2} \frac{\mathrm{d}Q_f}{\mathrm{d}T_e} \right)  \int_{E_\nu^\text{min}}^{E_\nu^\text{max}} \hspace{-0.5cm}\mathrm{d}E_\nu \frac{\mathrm{d}\Phi_{\bar{\nu}_e}}{\mathrm{d}E_\nu} \left[\frac{\mathrm{d}\sigma_{\bar{\nu}_e}}{\mathrm{d}T_\mathcal{N}} \right]_\xi^{\mathrm{CE}\nu\mathrm{NS}}\, ,
\end{equation}    
\end{widetext}
where $\xi = \{\text{SM, BSM}\}$ specifies the interaction type under consideration. In this expression, $t_{\text{run}}$ is the total data acquisition time, while $N_{\text{target}}$ represents the number of target nuclei in the detector, given by $N_{\text{target}} = m_{\text{det}} N_A/m_{\text{Ge}}$. Here, $m_{\text{det}}$ is the detector mass, $N_A$ is Avogadro number, and $m_{\text{Ge}}$ is the molar mass of the germanium isotopes. We consider all stable isotopes of Ge, including $\{^{70}\mathrm{Ge}, ^{72}\mathrm{Ge}, ^{73}\mathrm{Ge}, ^{74}\mathrm{Ge}, ^{76}\mathrm{Ge}\}$ with their respective natural abundances $\{20.57, 27.45, 7.75, 36.50, 7.73\}\%$~\cite{BerglundWieser:2011}. The minimum neutrino energy required for producing a nuclear recoil of energy $T_\mathcal{N}$ is determined by the kinematics of the interaction, $E_\nu^\text{min} \approx \sqrt{m_\mathcal{N} T_\mathcal{N}/2}$, while $E_\nu^\text{max}$ is the maximum neutrino energy of the reactor antineutrino spectrum. The electron-equivalent ionization energy, $T_e$, is related to the nuclear recoil energy, $T_\mathcal{N}$, via the quenching factor $Q_f$, which quantifies the fraction of nuclear recoil energy that produces ionization. This relationship is expressed as $T_e = Q_f(T_\mathcal{N}) \cdot T_\mathcal{N}$. For the CONUS+ experiment, the quenching factor is modeled using Lindhard theory~\cite{Lindhard:1963} with the parameter $k = 0.162$, as reported by the collaboration~\cite{Ackermann:2025obx, Bonhomme:2022lcz}. 
Finally, the event spectrum undergoes smearing through a Gaussian resolution function, given by,
\begin{equation}
    \mathcal{G}\left(T_e,T_e^\mathrm{reco}\right) = \frac{1}{\sqrt{2\pi}~\sigma(T_e)}  \exp{\left(-\left[\frac{T_e^\mathrm{reco} - T_e}{\sqrt{2}~\sigma(T_e)}\right]^2 \right)}\,,
\end{equation}
where $T_e^\mathrm{reco}$ represents the reconstructed (or measured) ionization energy. The energy-dependent resolution of the detector is modeled as $\sigma(T_e) = \sqrt{\sigma_0^2 + \eta F_f T_e}$~\cite{Bonhomme:2022lcz, Ackermann:2025obx}. Here, $\eta = 2.96 \mathrm{~eV}_{ee}$ denotes the mean energy required to create an electron-hole pair in germanium at $90$ K, while the Fano factor for germanium is set to $F_f = 0.1096$~\cite{Ackermann:2025obx}. The term $\sigma_0$ represents the peak resolution of the detector, determined from artificial signals produced using a pulse generator (Keysight 33500B) with a rise time matching that of physical signals. The collaboration has reported the measured pulser resolution in terms of full-width at half-maximum (FWHM)~\cite{CONUS:2024lnu, Ackermann:2025obx}, which relates to $\sigma_0$ for a Gaussian distribution as $\sigma_0 = \mathrm{FWHM}/(2\sqrt{2\ln{2}})$. For the measured value of $\mathrm{FWHM} = 48 \mathrm{~eV}_{ee}$, we obtain $\sigma_0 = 20.38 \mathrm{~eV}_{ee}$. Ultimately, the differential event rate is integrated over the electron-equivalent ionization energy $T_e$, ranging from $T_e^\mathrm{min} = 2.96 \mathrm{~eV}_{ee}$, which corresponds to the minimum energy required to generate an electron-hole pair in Ge, up to $T_e^\mathrm{max}$, corresponding to the maximum ionization energy as allowed by the kinematics.

The flux distribution of reactor antineutrinos, $\bar{\nu}_e$, is expressed as:  
\begin{equation}
\label{Eq:Reactor_Flux}
\frac{\mathrm{d}\Phi_{\bar{\nu}_e}}{\mathrm{d}E_\nu} = \frac{P}{4 \pi d^2 \epsilon} \, \sum_{k} f_k\frac{\mathrm{d} N^k_{\bar{\nu}_e}}{\mathrm{d} E_\nu}\, ,
\end{equation}
where $P$ denotes the thermal power of the reactor, $d$ is the distance from the reactor core to the detector, and $\epsilon$ represents the average energy released per fission. The term $\mathrm{d} N^k_{\bar{\nu}_e}/\mathrm{d} E_\nu$ corresponds to the energy spectrum of antineutrinos emitted by different production mechanisms, indexed by $k$.  Reactor antineutrinos primarily originate from two key sources: (a) the beta decays of fission fragments from the isotopes $\mathrm{^{235}U}$, $\mathrm{^{238}U}$, $\mathrm{^{239}Pu}$, and $\mathrm{^{241}Pu}$. For these contributions, the antineutrino spectra are modeled using the Huber-Müller parametrization~\cite{Huber:2011wv, Mueller:2011nm} above 2 MeV and from Ref.~\cite{Vogel:1989iv} for lower energies. As reported by the CONUS+ collaboration in Ref.~\cite{Ackermann:2025obx}, during reactor operation, the average fission fractions ($f_k$) of these fissile isotopes are considered to be 53\%, 8\%, 32\%, and 7\%, respectively;   (b) neutron capture on $\mathrm{^{238}U}$, leading to the production of $\mathrm{^{239}U}$ ($\mathrm{^{238}U(n, \gamma)^{239}U}$). The spectrum of this process is adopted from Ref.~\cite{TEXONO:2006xds}. The total antineutrino flux is obtained by summing over all relevant contributions: $k = \{\mathrm{^{235}U}, \mathrm{^{238}U}, \mathrm{^{239}Pu}, \mathrm{^{241}Pu}, \mathrm{^{238}U(n, \gamma)^{239}U}\}$, weighted by their respective fission fractions. For the CONUS+ experiment, the experimental baseline is $d = 20.7$ m, the reactor thermal power is $P = 3.6$ GW, and the average energy release in each fission is $\epsilon = 205.24$ MeV.  Using these the total flux normalization at the detector is estimated to be $\mathscr{N} = 1.5 \times 10^{13} \mathrm{~cm}^{-2} \mathrm{s}^{-1}$~\cite{Ackermann:2025obx}.

On the other hand, the predicted number of theoretical \eves events in the $i$th electron-recoil energy bin is determined as:  
\begin{equation}
\label{equn:EvES_Events}
\left[R_i\right]_\xi^{\text{E}\nu\text{ES}} = t_{\text{run}} N_{\text{target}} \int_{T_e^i}^{T_e^{i+1}} \hspace{-0.7cm}\mathrm{d}T_e^\mathrm{reco}\hspace{0.05cm}\int_{T_e^\mathrm{min}}^{T_e^\mathrm{max}}\hspace{-0.6cm}\mathrm{d}T_e~\mathcal{G}\left(T_e,T_e^\mathrm{reco}\right) Z_\mathrm{eff}(T_e)   \int_{E_\nu^\text{min}}^{E_\nu^\text{max}} \hspace{-0.5cm}\mathrm{d}E_\nu \frac{\mathrm{d}\Phi_{\bar{\nu}_e}}{\mathrm{d}E_\nu} \left[\frac{\mathrm{d}\sigma_{\bar{\nu}_e}}{\mathrm{d}T_e} \right]_\xi^{\mathrm{E}\nu\mathrm{ES}}\, .
\end{equation}
For \eves the kinematics of the process gives $E_\nu^\mathrm{min}=\left[T_e+\sqrt{T_e^2+2m_eT_e}\right]/2$. The effective number of electrons, $Z_\mathrm{eff}(T_e)$, available for ionization at a given recoil energy $T_e$, incorporates the effects of atomic binding. This quantity is approximated via a series of step functions, as discussed in Ref.~\cite{Chen:2016eab}, and is expressed as $Z_\mathrm{eff} (T_e)=\sum_{j=1}^{32}\Theta(T_e-\mathscr{B}_j)$, where $\Theta(x)$ is the Heaviside step function, and $\mathscr{B}_j$ represents the binding energy of the $j$th electron in a germanium atom. The binding energy values are taken from the data presented in Ref.~\cite{xray_data_booklet}. This approximation accounts for the energy thresholds of each atomic electron, ensuring an accurate representation of the ionization process within the detector.

\begin{figure}[ht!]
    \centering
    \includegraphics[width=0.49\linewidth]{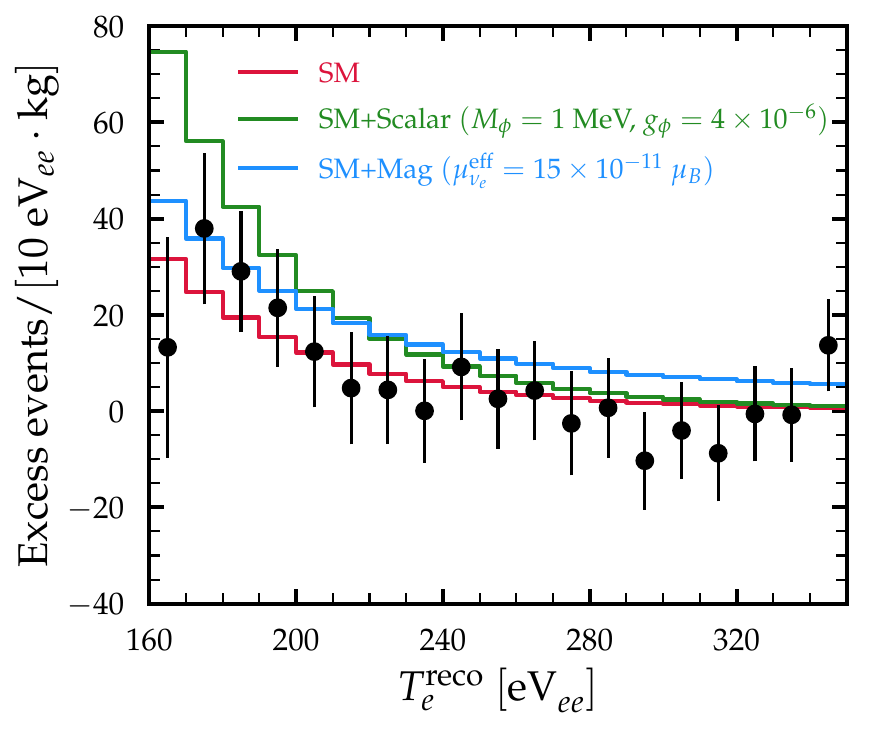}
    \caption{Simulated signals (colored histograms) and background-subtracted reactor-on data (black points with error bars) as a function of $T_e^\mathrm{reco}$ at the CONUS+ experiment, with an experimental run time of 119 days. The simulated signals include contributions from both \cevns and \eves events, as considered in this work, and are computed for an effective detector configuration with a 1 kg target mass and a threshold of $160 \mathrm{~eV}_{ee}$, following the prescription of the CONUS+ collaboration.}
    \label{fig:Events}
\end{figure}

Figure~\ref{fig:Events} illustrates the simulated events and the background-subtracted reactor-on data observed at the CONUS+ experiment over an experimental run time of $119$ days, plotted as a function of $T_e$. The simulated signals, shown as colored histograms, represent the predicted events spectra for the SM as well as various new physics scenarios. In all cases, the simulated spectra include contributions from both \cevns and E$\nu$ES. Notably, to compare with the excess event data reported by the CONUS+ collaboration~\cite{Ackermann:2025obx}, we compute the predicted events considering a single effective detector with a mass of 1 kg, a threshold of $160 \mathrm{~eV}_{ee}$, and an experimental run time of $119$ days, as prescribed by the CONUS+ collaboration (private communication).

The statistical analysis employed in this study is based on the Gaussian $\chi^2$ function, which is given by~\cite{Almeida:1999ie}:
\begin{widetext}
\begin{equation}
\label{equn:chi_square}
\chi^2(\overrightarrow{S}; \alpha) = \sum_{i} \left[ \frac{(1 + \alpha) R_i^{\text{CE}\nu\text{NS}} (\overrightarrow{S}) + (1 + \beta)R_i^{\text{E}\nu\text{ES}} (\overrightarrow{S}) - R_i^\text{exp}}{\sigma_i^\text{exp}} \right]^2 + \left( \frac{\alpha}{\sigma_\alpha} \right)^2 + \left( \frac{\beta}{\sigma_\beta} \right)^2 \, ,
\end{equation}    
\end{widetext}
where $R_i^{\text{CE}\nu\text{NS}}$ and $R_i^{\text{E}\nu\text{ES}}$ represent the theoretical predictions for the \cevns and \eves events in the $i$th bin, respectively. $\overrightarrow{S}$ contains the set of free parameters that we aim to determine. Additionally, $R_i^\text{exp}$ and $\sigma_i^\text{exp}$ denote the experimentally observed number of residual events (background-subtracted reactor-on data) and their associated statistical uncertainties in the $i$th bin, as provided in the data release~\cite{Ackermann:2025obx}.  
The cumulative systematic uncertainties in the predicted \cevns and \eves events, accounting for uncertainties associated with the reactor antineutrino flux spectrum estimation (4.6\%), quenching effects (7.3\%), energy threshold (14.1\%), nuclear form factor (3.2\%), the active Ge mass (1.1\%), and trigger efficiency (0.7\%), are incorporated via the nuisance parameters $\alpha$ and $\beta$ for \cevns and \eves events, respectively, with their associated uncertainties, $\sigma_\alpha = 16.89\%$ and $\sigma_\beta = 14.89\%$~\cite{Ackermann:2025obx}. Notably, the uncertainties associated with the quenching factor and nuclear form factor are considered only for \cevns events and are not applied to \eves events.
For the analysis of a given parameter from the set $\overrightarrow{S}$, we perform marginalization over the nuisance parameters $\alpha$ and $\beta$ in the $\chi^2$ function. 

\FloatBarrier
\section{Results}
\label{Sec:Results}

\begin{figure}[h]
    \centering
    \includegraphics[width=0.45\linewidth]{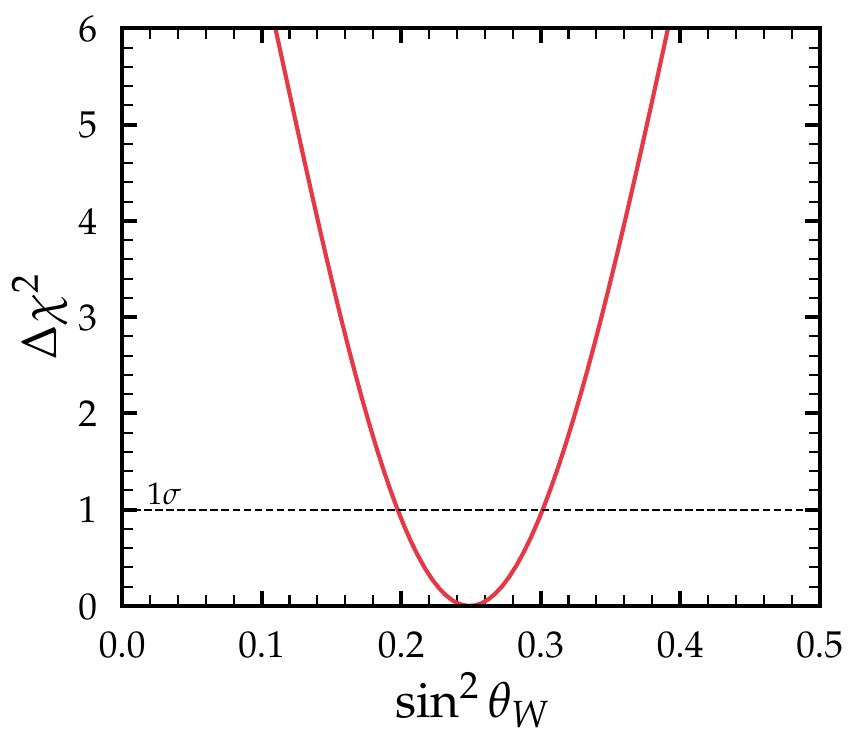}
    \caption{$\Delta\chi^2$ profile of weak mixing angle obtained from CONUS+ data.}
    \label{fig:sW2_chi2}
\end{figure}

\begin{figure*}[ht!]
    \centering
    \includegraphics[width=0.8\linewidth]{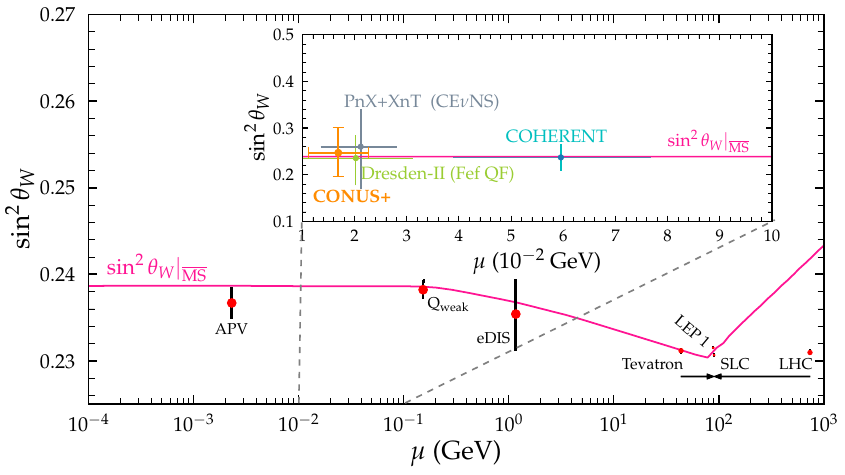}
    \caption{Comparison of the weak mixing angle results from CONUS+ with other experimental constraints at different renormalization scales.}
    \label{fig:sW2_Comparison}
\end{figure*}

In this section, we present the sensitivities derived from the CONUS+ experiment for various physics scenarios discussed in Sec.~\ref{Sec:Theory}. We begin by examining the constraints on the weak mixing angle. Figure~\ref{fig:sW2_chi2} shows the $\Delta\chi^2$ profile for $\sin^2{\theta_W}$. 
From this, we extract the $1\sigma$ determination of the weak mixing angle as:
\begin{equation*}
    \sin^2{\theta_W} = 0.247^{+0.050}_{-0.054}.
\end{equation*}
In Fig.~\ref{fig:sW2_Comparison}, we compare our results, derived from the analysis of CONUS+ data, with other determinations from experiments spanning a wide energy range~\cite{Majumdar:2022nby, DeRomeri:2022twg, DeRomeri:2024iaw, ParticleDataGroup:2024cfk}.
It is important to note that while the CONUS+ constraint is comparable to that from the Dresden-II experiment~\cite{Majumdar:2022nby} using the FeF quenching factor, it is less stringent than the limits obtained from COHERENT CsI+LAr data~\cite{DeRomeri:2022twg}. However, the CONUS+ result demonstrates a significantly higher precision compared to the limits derived from recent measurements of solar $^8\mathrm{B}$ neutrino-induced \cevns signals by PandaX-4T and XENONnT~\cite{DeRomeri:2024iaw}. Finally, our determination of $ \sin^2{\theta_W}$ is also consistent with that obtained in the recent Ref.~\cite{Alpizar-Venegas:2025wor} which appeared on arXiv while this manuscript was under preparation.

\begin{figure}[ht!]
    \centering
    \includegraphics[width=0.45\linewidth]{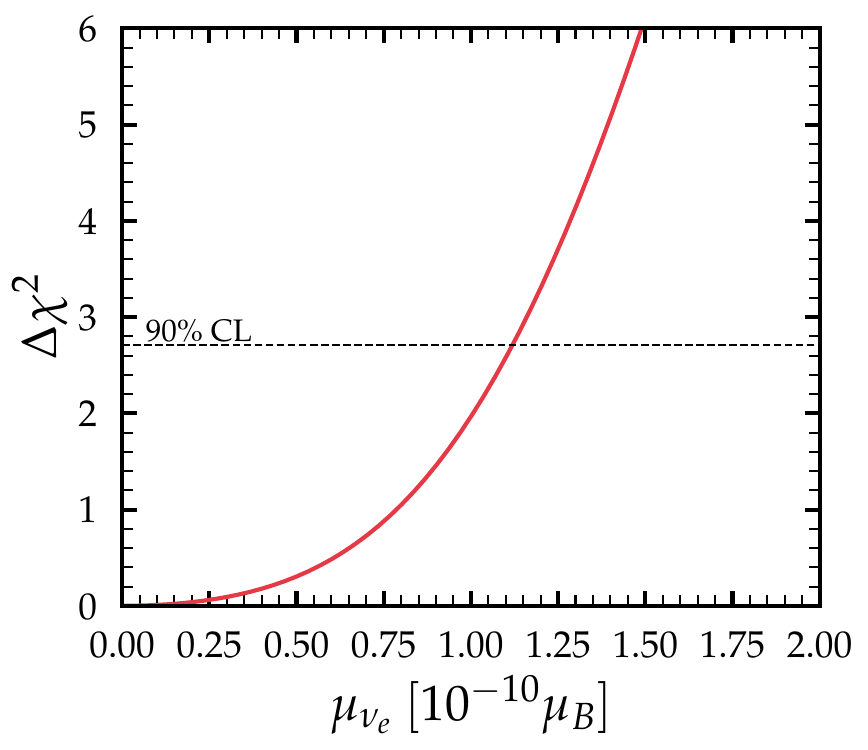}
    \caption{$\Delta\chi^2$ profile of effective neutrino magnetic moment obtained from CONUS+ data exploiting both \cevns and \eves signals.}
    \label{fig:magnetic_moment}
\end{figure}

\begin{table}[ht!]
    \centering
    \begin{tabular}{|@{\hspace{0.5cm}}c@{\hspace{0.5cm}}|}
    \hline
        $\mathbf{\mu^\text{eff}_{\nu_e}~ (10^{-11}~\mu_B)}$\\
        \hline
        \hline
        $\mathbf{\leq11.2}$~\textbf{(CONUS+)}\\
        $\leq7.5$~(CONUS)~\cite{CONUS:2022qbb}\\
        $\leq420$~(COHERENT CsI+LAr)~\cite{AtzoriCorona:2022qrf}\\
        $\leq20.8$~(DRESDEN-II)~\cite{AtzoriCorona:2022qrf}\\
        $\leq3.9$~(Borexino)~\cite{Borexino:2017fbd, Coloma:2022umy}\\
        $\leq7.4$~(TEXONO)~\cite{TEXONO:2006xds}\\
        $\leq2.9$~(GEMMA)~\cite{Beda:2012zz}\\
        $\leq1.4$~(LZ)~\cite{A:2022acy}\\
        $\leq0.9$~(XENONnT)~\cite{A:2022acy}\\
         \hline
    \end{tabular}
    \caption{Comparison of 90\% C.L. upper limits on the effective neutrino magnetic moment from various experiments.}
    \label{tab:magnetic_moment_limit}
\end{table}

We now turn our attention to the constraints on the neutrino magnetic moment obtained from the CONUS+ experiment, utilizing both \cevns and \eves signals. Since reactor neutrinos primarily constitute the $\bar{\nu}_e$ flux, the only relevant parameter in this analysis is $\mu_{\nu_e}^\mathrm{eff}$. The $\Delta\chi^2$ profile for $\mu_{\nu_e}^\mathrm{eff}$ is shown in Fig.~\ref{fig:magnetic_moment}. By combining the \cevns and \eves signals, we obtain the upper limit at 90\% C.L. as $\mu_{\nu_e}^\mathrm{eff} \leq 1.12 \times 10^{-10}~\mu_B$. Using only the \cevns signal, we find the upper limit to be $\mu_{\nu_e}^\mathrm{eff} \leq 4.76 \times 10^{-10}~\mu_B$. Clearly, incorporating \eves along with \cevns signals results in a more stringent limit. Notably, the limit obtained from the CE$\nu$NS-only analysis in this study is consistent with the results reported in the recent Ref.~\cite{Alpizar-Venegas:2025wor}. In Tab.~\ref{tab:magnetic_moment_limit}, we compare the 90\% C.L. upper limits on $\mu_{\nu_e}^\mathrm{eff}$ from CONUS+ with those from other experiments. As can be seen, while the CONUS+ limit is more precise than those from COHERENT and Dresden-II experiments, it is still less stringent than the limits obtained from E$\nu$ES-based analyses of solar neutrino data, such as those from XENONnT, LZ, and Borexino.

\begin{figure*}[ht!]
    \centering
    \includegraphics[width=0.49\linewidth]{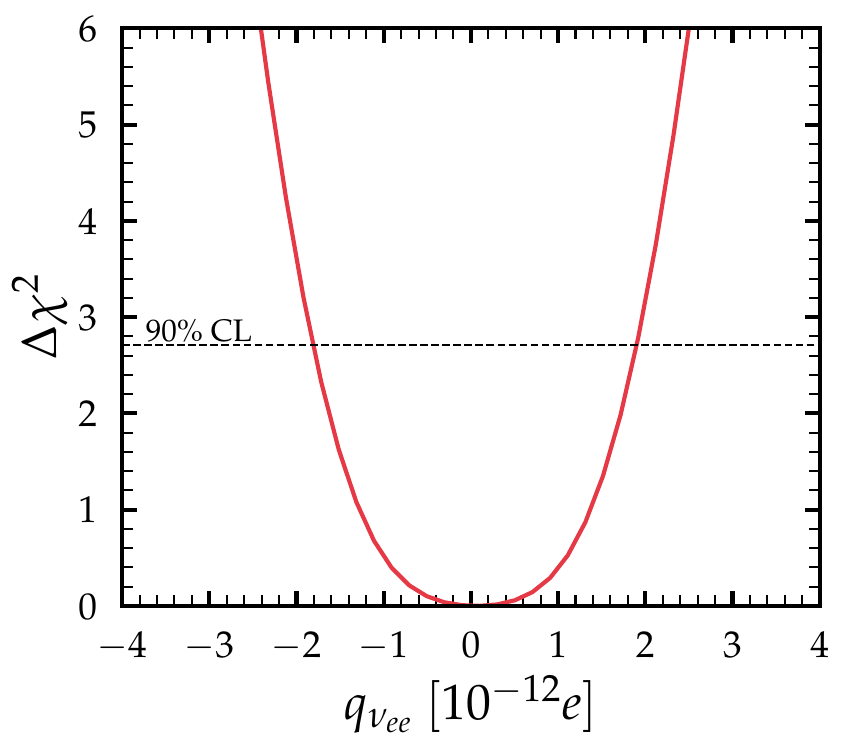}
    \includegraphics[width=0.49\linewidth]{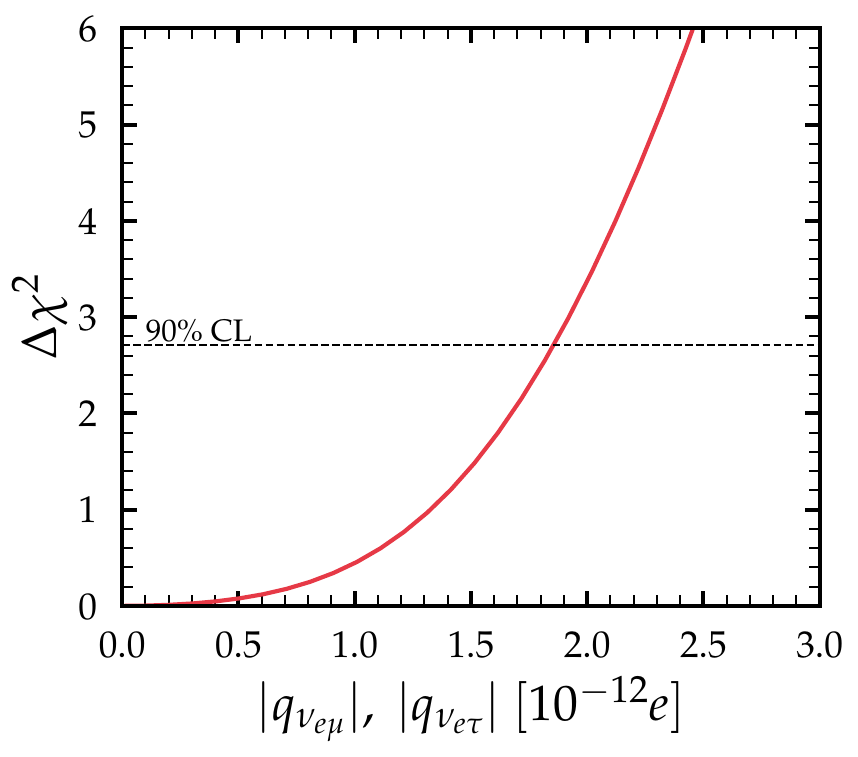}
    \caption{Marginalized $\Delta\chi^2$ profiles for $q_{\nu_{ee}}$ (left) and $q_{\nu_{e\mu}}$, $q_{\nu_{e\tau}}$ (right) obtained from CONUS+ data exploiting both \cevns and \eves signals. Notably, the limits on $q_{\nu_{e\mu}}$ and $q_{\nu_{e\tau}}$ are same (see text for details).}
    \label{fig:millicharge}
\end{figure*}

\begin{table*}[ht!]
    \centering
    \begin{tabular}{|@{\hspace{0.06cm}}c@{\hspace{0.06cm}}|@{\hspace{0.06cm}}c@{\hspace{0.06cm}}|@{\hspace{0.06cm}}c@{\hspace{0.06cm}}|}
    \hline
        $\mathbf{q_{\nu_{ee}}~ (10^{-12}~e)}$ &  $\mathbf{|q_{\nu_{e\mu}}|~ (10^{-12}~e)}$ & $\mathbf{|q_{\nu_{e\tau}}|~ (10^{-12}~e)}$ \\
        \hline
        \hline
        $\mathbf{[-1.8,1.9]}$~\textbf{(CONUS+)} & $\mathbf{\leq1.85}$~\textbf{(CONUS+)} & $\mathbf{\leq1.85}$~\textbf{(CONUS+)} \\
        $\leq3.3$~(CONUS)~\cite{CONUS:2022qbb} &  &  \\
         $[-500,500]$~(COHERENT CsI+LAr)~\cite{AtzoriCorona:2022qrf} & $\leq180$~(COHERENT CsI+LAr)~\cite{AtzoriCorona:2022qrf} & $\leq500$~(COHERENT CsI+LAr)~\cite{AtzoriCorona:2022qrf} \\
         $[-8.6,8.7]$~(DRESDEN-II)~\cite{AtzoriCorona:2022qrf} & $\leq8.6$~(DRESDEN-II)~\cite{AtzoriCorona:2022qrf} & $\leq8.6$~(DRESDEN-II)~\cite{AtzoriCorona:2022qrf} \\
         $[-0.3,0.6]$~(LZ)~\cite{A:2022acy}  &  &  \\
         $[-0.1,0.6]$~(XENONnT)~\cite{A:2022acy}  &  &  \\
         \hline
    \end{tabular}
    \caption{Comparison of 90\% C.L. limits on the neutrino millicharge from various experiments.}
    \label{tab:millicharrge_limit}
\end{table*}

We now discuss the constraints on the neutrino millicharge obtained from the CONUS+ experiment. The relevant parameters in this analysis are $q_{\nu_{ee}}$, $q_{\nu_{e\mu}}$, and $q_{\nu_{e\tau}}$. During the fitting process, we allow the remaining two parameters to vary freely in order to compute the theoretical events. For each parameter of interest, we then marginalize the $\chi^2$ function over the other two parameters. Further, one should note that as reactor neutrinos consist solely of $\bar{\nu}_e$ flux, and the cross section in Eq.~\eqref{Eq:millichare_CR} is symmetric with respect to $q_{\nu_{e\mu}}$ and $q_{\nu_{e\tau}}$, the obtained limits on these two parameters from CONUS+ data are same.  Figure~\ref{fig:millicharge} displays the marginalized $\Delta\chi^2$ profiles for $q_{\nu_{ee}}$, $q_{\nu_{e\mu}}$, and $q_{\nu_{e\tau}}$ obtained by analyzing both \cevns and \eves signals. At 90\% C.L., the limits are:
\begin{equation*}
    \{q_{\nu_{ee}}, |q_{\nu_{e\mu}}|, |q_{\nu_{e\tau}}|\} = \{[-1.8,1.9], (\leq 1.85), (\leq 1.85)\} \times 10^{-12}~e.
\end{equation*}
For comparison, using only the \cevns signal, the 90\% C.L. limits are:
\begin{equation*}
    \{q_{\nu_{ee}}, |q_{\nu_{e\mu}}|, |q_{\nu_{e\tau}}|\} = \{[-0.49,3.22], (\leq 1.9), (\leq 1.9)\} \times 10^{-8}~e.
\end{equation*}
Clearly, the inclusion of the \eves signal alongside the \cevns signal significantly improves the constraints on the neutrino millicharge. Table~\ref{tab:millicharrge_limit} summarizes the 90\% C.L. limits from CONUS+ and compares them with results from other experiments. As shown, the CONUS+ constraints are more stringent than those from COHERENT and Dresden-II experiments. However, they remain less stringent than limits derived from E$\nu$ES-based analyses of solar neutrino data, such as those from XENONnT and LZ.

\begin{figure*}[ht!]
    \centering
    \includegraphics[width=0.49\linewidth]{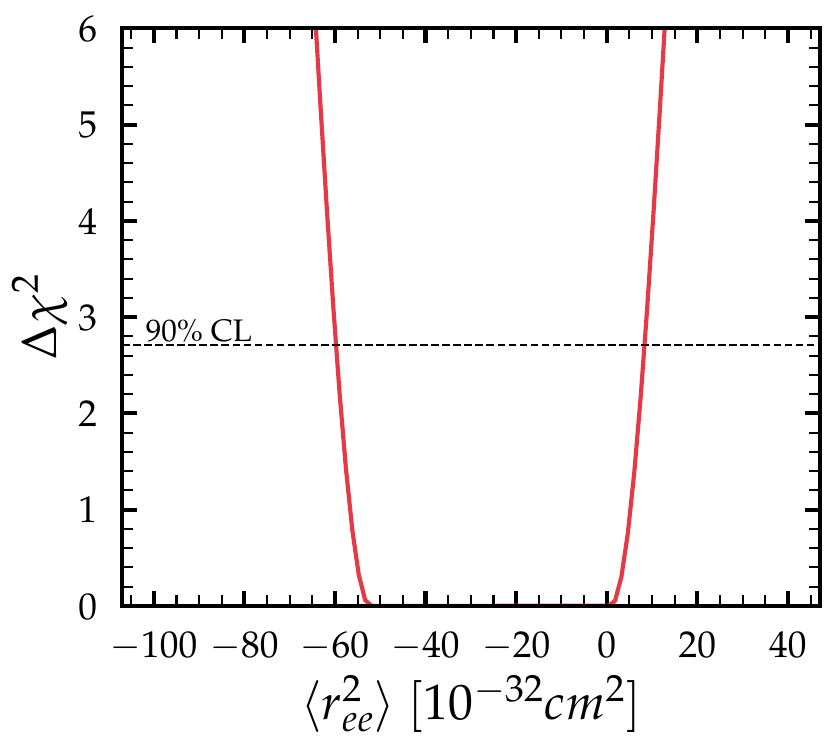}
    \includegraphics[width=0.49\linewidth]{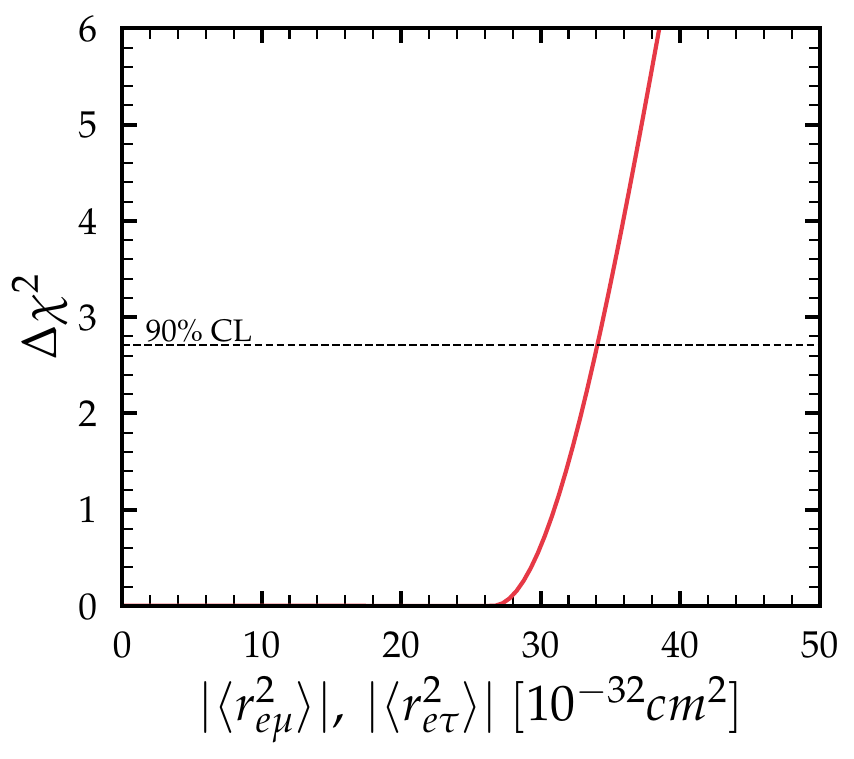}
    \caption{Marginalized $\Delta\chi^2$ profiles for $\langle r^2_{\nu_{ee}}\rangle$ (left) and $\langle r^2_{\nu_{e\mu}}\rangle$, $\langle r^2_{\nu_{e\tau}}\rangle$ (right) obtained from CONUS+ data, exploiting both \cevns and \eves signals. Notably, the limits on $\langle r^2_{\nu_{e\mu}}\rangle$ and $\langle r^2_{\nu_{e\tau}}\rangle$ are same (see text for details).}
    \label{fig:CR}
\end{figure*}

\begin{table*}[ht!]
    \centering
    \begin{tabular}{|@{\hspace{0.1cm}}c@{\hspace{0.1cm}}|@{\hspace{0.1cm}}c@{\hspace{0.1cm}}|@{\hspace{0.1cm}}c@{\hspace{0.1cm}}|}
    \hline
        $\mathbf{\langle r^2_{\nu_{ee}}\rangle~ (10^{-32}~\mathrm{cm}^2)}$ &  $\mathbf{\langle r^2_{\nu_{e\mu}}\rangle~ (10^{-32}~\mathrm{cm}^2)}$ & $\mathbf{\langle r^2_{\nu_{e\tau}}\rangle~ (10^{-32}~\mathrm{cm}^2)}$\\
        \hline
        \hline
        $\mathbf{[-59.76,8.33]}$~\textbf{(CONUS+)} & $\mathbf{\leq34.06}$~\textbf{(CONUS+)} & $\mathbf{\leq34.06}$~\textbf{(CONUS+)}\\
         $[-57,4]$~(DRESDEN-II)~\cite{AtzoriCorona:2022qrf} & $\leq30$~(DRESDEN-II)~\cite{AtzoriCorona:2022qrf} & $\leq30$~(DRESDEN-II)~\cite{AtzoriCorona:2022qrf}\\
         $[-69,14]$~(COHERENT CsI+LAr)~\cite{AtzoriCorona:2022qrf,Khan:2022akj} & $\leq30$~(COHERENT CsI+LAr)~\cite{AtzoriCorona:2022qrf} & $\leq42$~(COHERENT CsI+LAr)~\cite{AtzoriCorona:2022qrf} \\
          $[-4.2,6.6]$~(TEXONO)~\cite{TEXONO:2009knm} &  & \\ $[-5.94,8.28]$~(LSND)~\cite{LSND:2001akn} &  &\\
           $[-121,37.5]$~(LZ)~\cite{A:2022acy} &  & \\
           $[-93.4,9.5]$~(XENONnT)~\cite{A:2022acy} &  &  \\
         \hline
    \end{tabular}
    \caption{Comparison of 90\% C.L. limits on the neutrino CR parameters from CONUS+ and other experiments.}
    \label{tab:CR_limit}
\end{table*}

Turning to neutrino charge radius (CR), the relevant neutrino CR parameters in this analysis are $\langle r^2_{\nu_{ee}}\rangle$, $\langle r^2_{\nu_{e\mu}}\rangle$, and $\langle r^2_{\nu_{e\tau}}\rangle$. Similar to the neutrino millicharge case, marginalized $\Delta\chi^2$ profiles for these parameters are shown in Fig.~\ref{fig:CR}. Furthermore, the obtained limits on the transition CR parameters $\langle r^2_{\nu_{e\mu}}\rangle$ and $\langle r^2_{\nu_{e\tau}}\rangle$ are identical in the CONUS+ analysis. At 90\% C.L., the limits obtained from CONUS+ are:
\begin{widetext}
\begin{equation*}
    \{\langle r^2_{\nu_{ee}}\rangle, |\langle r^2_{\nu_{e\mu}}\rangle|, |\langle r^2_{\nu_{e\tau}}\rangle|\} = \{[-59.76,8.33], (\leq 34.06), (\leq 34.06)\} \times 10^{-32}~\mathrm{cm}^2.
\end{equation*}   
\end{widetext}
It is worth noting that the contribution of \eves to the neutrino CR cross section is negligible. Therefore, including \eves data alongside \cevns signal does not effectively improve the constraints, yielding results consistent with a CE$\nu$NS-only analysis. Tab.~\ref{tab:CR_limit} compares the 90\% C.L. limits obtained from CONUS+ with those from other experiments. Notably, the CONUS+ limits on neutrino CR parameters are competitive with those obtained from DRESDEN-II and COHERENT experiments, they remain less stringent compared to constraints from TEXONO and LSND.

\begin{figure*}[ht!]
    \centering
    \includegraphics[width=0.49\linewidth]{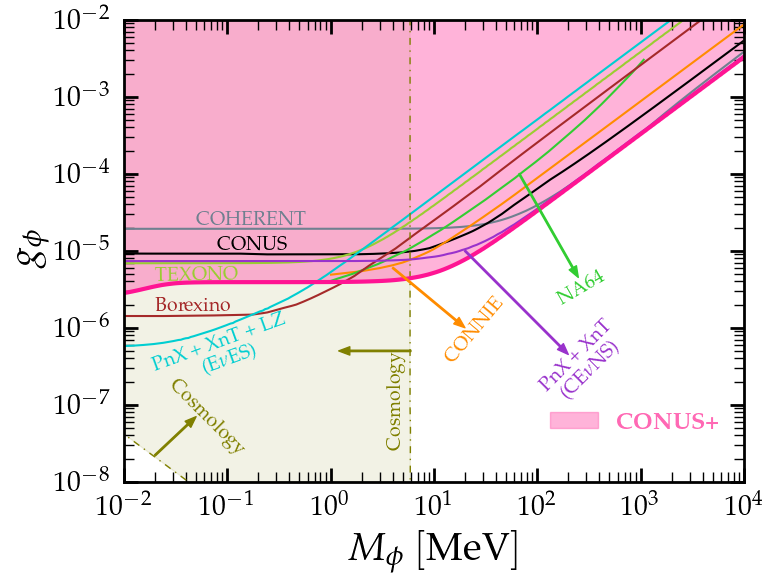}
    \includegraphics[width=0.49\linewidth]{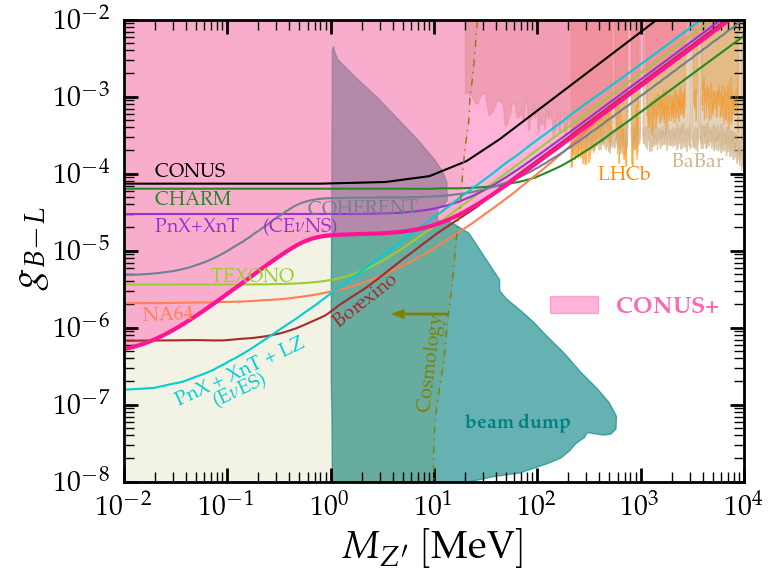}
    \caption{Exclusion limits at 90\% C.L. in the $(M_X, g_X)$ parameter space for light scalar and vector $B-L$ mediators derived from CONUS+ data, exploiting both \cevns and \eves signals. Existing experimental and astrophysical constraints are also shown for comparison, providing a comprehensive perspective on the probed parameter space (see text for details).}
    \label{fig:Light_Mediators_Param_Space}
\end{figure*}

We now present the constraints on light scalar and vector $B-L$ mediators derived from the CONUS+ experiment. Figure~\ref{fig:Light_Mediators_Param_Space} illustrates the 90\% C.L. exclusion contours in the $(M_X, g_X)$ parameter space. The inclusion of \eves events alongside \cevns significantly improves the constraints for vector $B-L$ mediators in the light mediator regime, particularly for $M_{Z'} \lesssim 1~\mathrm{MeV}$, while for scalar mediators, its inclusion leads to a slight improvement in the constraints for $M_\phi \lesssim 30~\mathrm{keV}$.

A comparison of these results with constraints from other experimental and astrophysical studies highlights their complementarity. In Fig.~\ref{fig:Light_Mediators_Param_Space}, we include exclusion limits derived from the combined analysis of CsI and LAr \cevns data of COHERENT collaboration~\cite{DeRomeri:2022twg, AtzoriCorona:2022moj, Majumdar:2024dms}. Notably, the CsI analysis incorporates both \cevns and \eves contributions, whereas the LAr results rely solely on \cevns events. Additionally, limits from existing CONUS data~\cite{CONUS:2021dwh}, as well as limits from recent measurements of solar $^8\mathrm{B}$ neutrino-induced \cevns signals by PandaX-4T and XENONnT~\cite{DeRomeri:2024iaw, Blanco-Mas:2024ale} are shown. The parameter space probed by CONUS+ is further compared with constraints from E$\nu$ES-bassed constraints, including BOREXINO~\cite{Coloma:2022avw}, CHARM-II~\cite{CHARM-II:1994wkf}, and TEXONO~\cite{Bilmis:2015lja, Majumdar:2024dms}, as well as combined analyses of data from PandaX-4T, XENONnT, and LZ~\cite{A:2022acy, DeRomeri:2024dbv, Majumdar:2024dms}. We also present limits from dark photon searches carried out at various fixed-target experiments, including beam-dump facilities such as CHARM~\cite{CHARM:1985anb,Gninenko:2012eq}, NA64~\cite{NA64:2016oww,NA64:2019auh, NA64:2023wbi}, NOMAD~\cite{NOMAD:2001eyx}, E141~\cite{Riordan:1987aw,Bjorken:2009mm}, E137~\cite{Bjorken:1988as,Andreas:2012mt}, E774~\cite{Bross:1989mp}, KEK~\cite{Konaka:1986cb}, Orsay~\cite{Andreas:2012mt}, U70/$\nu$-CAL~I~\cite{Blumlein:2011mv,Blumlein:2013cua}, and APEX~\cite{APEX:2011dww}. High-energy collider experiments, such as BaBar~\cite{BaBar:2014zli,BaBar:2017tiz} and LHCb~\cite{LHCb:2019vmc}, have also set constraints on dark photon models. These dark photon search limits have been recasted into the relevant parameter space using the darkcast software, as detailed in Refs.~\cite{Ilten:2018crw, Baruch:2022esd}. Astrophysical and cosmological observations impose further limits. Constraints from cosmological bounds on the relativistic degrees of freedom ($N_\mathrm{eff}$)~\cite{Esseili:2023ldf, Li:2023puz, Ghosh:2024cxi}, as well as Big Bang Nucleosynthesis (BBN) limits~\cite{Blinov:2019gcj, Suliga:2020jfa}, are also included. 

The complementarity of the CONUS+ constraints on scalar and vector $B-L$ mediators is particularly evident in Fig.~\ref{fig:Light_Mediators_Param_Space}. Notably, for scalar mediators, the CONUS+ bounds provide the most stringent constraint for $M_\phi > 6~\mathrm{MeV}$. Focusing exclusively on constraints derived from CE$\nu$NS data, the CONUS+ results surpass all existing CE$\nu$NS-based limits for both scalar and vector $B-L$ interactions, except for $M_{Z'} \gtrsim 100~\mathrm{MeV}$ in the vector $B-L$ case, where the COHERENT bound is more stringent. 

\FloatBarrier
\section{Conclusions}
\label{Sec:Conclusions}

The CONUS+ experiment has recently reported the reactor antineutrino induced \cevns data. We have analyzed the SM and BSM physics implications of this data set. Within the framework of the SM, we looked at the estimation of weak mixing angle and found that the constraint derived from CONUS+ data exhibit a significantly improved precision compared to the limits from recent measurements of solar $^8\mathrm{B}$ neutrino-induced \cevns signals by PandaX-4T and XENONnT. The CONUS+ limit is also comparable to those from the Dresden-II experiment using the FeF quenching factor. However, it is less stringent than the limits obtained from COHERENT CsI+LAr data. For BSM physics we looked at several ascpects such as neutrino EM properties as well as NGI with light mediators. The CONUS+ limits on neutrino electromagnetic (EM) properties, including the charge radius, millicharge, and magnetic moment, are comparable to those extracted from the COHERENT and Dresden-II data, and in some cases, provide improved constraints. Notably, the inclusion of \eves signals alongside the \cevns data significantly enhances the constraints on the neutrino magnetic moment and millicharge. However, the obtained CONUS+ limits remain less stringent compared to constraints derived from E$\nu$ES-based experiments, such as TEXONO, Borexino, LZ, and XENONnT. Regarding NGI with light mediators, we analyzed scalar and vector $B-L$ interaction scenarios. The CONUS+ results provide complementary constraints in these scenarios. Specifically, for scalar mediators, the CONUS+ bounds are the most stringent for $M_\phi > 6~\mathrm{MeV}$. Focusing exclusively on CE$\nu$NS-based constraints, the CONUS+ results provide the strongest limits on scalar mediators across the entire parameter space and on vector $B-L$ mediators for $M_{Z'} \lesssim 100~\mathrm{MeV}$. This underscores the importance of the CONUS+ data in probing physics beyond the SM.

\acknowledgments
The authors are indebted to Dimitris Papoulias for his insightful comments which have been invaluable in completion of this work.
The authors also acknowledge Christian Buck for valuable correspondence and assistance regarding the interpretation of the data release from the CONUS+ collaboration.
AM expresses sincere thanks for the financial support provided through the Prime Minister Research Fellowship (PMRF), funded by the Government of India (PMRF ID: 0401970).

\bibliographystyle{utphys}
\bibliography{bibliography}
\end{document}